\documentclass[twocolumn]{aastex631}

\usepackage{mathrsfs}
\usepackage{rotating}
\usepackage{verbatim}
\usepackage[T1]{fontenc} 

\newcommand{\code}[1]{{\texttt{#1}}}

\begin{document}

\title{Short-Period Small Planets with High Mutual Inclinations are more Common around Metal-Rich Stars}

\correspondingauthor{Xinyan Hua, Sharon Xuesong Wang}
\email{huaxinyan1996@gmail.com, sharonw@tsinghua.edu.cn}

\author[0000-0001-7916-4371]{Xinyan Hua}
\affiliation{Department of Astronomy, Tsinghua University, Beijing 100084, China}

\author[0000-0002-6937-9034]{Sharon Xuesong Wang} 
\affiliation{Department of Astronomy, Tsinghua University, Beijing 100084, China}

\author[0000-0002-4847-3929]{Dongsheng An}
\affiliation{School of Astronomy and Space Science, Nanjing University, Nanjing 210023, China}
\affiliation{Key Laboratory of Modern Astronomy and Astrophysics, Ministry of Education, Nanjing 210023, China}
\author[0000-0002-7846-6981]{Songhu Wang} 
\affiliation{Department of Astronomy, Indiana University, 727 East 3rd Street, Bloomington, IN 47405-7105, USA}
\author[0000-0003-3250-2876]{Yang Huang}
\affiliation{School of Astronomy and Space Science, University of Chinese Academy of Sciences, Beijing 100049, China;}
\affiliation{Key Lab of Optical Astronomy, National Astronomical Observatories, Chinese Academy of Sciences, Beijing 100101, China}
\author[0000-0003-0707-3213]{Dichang Chen}
\affiliation{School of Astronomy and Space Science, Nanjing University, Nanjing 210023, China}
\affiliation{Key Laboratory of Modern Astronomy and Astrophysics, Ministry of Education, Nanjing 210023, China}
\author[0000-0003-0426-6634]{Johannes Buchner}
\affil{Max-Planck-Institut für extraterrestrische Physik, Giessenbachstrasse 1, D-85748 Garching, Germany}
\author[0000-0003-4027-4711]{Wei Zhu} 
\affiliation{Department of Astronomy, Tsinghua University, Beijing 100084, China}
\author[0000-0002-8958-0683]{Fei Dai}
\affiliation{Institute for Astronomy, University of Hawai`i, 2680 Woodlawn Drive, Honolulu, HI 96822, USA}
\author[0000-0002-6472-5348]{Jiwei Xie}
\affiliation{School of Astronomy and Space Science, Nanjing University, Nanjing 210023, China}
\affiliation{Key Laboratory of Modern Astronomy and Astrophysics, Ministry of Education, Nanjing 210023, China}



\begin{abstract}
We present a correlation between the stellar metallicities and the mutual inclinations of multi-planet systems hosting short-period small planets ($a/R_{\star}<12$, $R_p< 4\ \rm R_{\oplus}$). We analyzed 89 multi-planet systems discovered by \textit{Kepler}, K2, and TESS, where the innermost planets have periods shorter than 10~days. We found that the mutual inclinations of the innermost two planets are higher and more diverse around metal-rich stars. The mutual inclinations are calculated as the absolute differences between the best-fit inclinations of the innermost two planets from transit modeling, which represent the lower limits of the true mutual inclinations. The mean and variance of the mutual inclination distribution of the metal-rich systems are $3.1^{\circ}$$\pm 0.5$ and $3.1^{\circ}$$\pm 0.4$, while for the metal-poor systems they are $1.3^{\circ}$$\pm 0.2$ and $1.0^{\circ}$$\pm 0.2$. This finding suggests that inner planetary systems around metal-rich stars are dynamically hotter. We summarized the theories that could plausibly explain this correlation, including the influence of giant planets, higher solid densities in protoplanetary disks around metal-rich stars, or secular chaos coupled with an excess of angular momentum deficits. Planet formation and population synthesis models tracking the mutual inclination evolution would be essential to fully understand this correlation. 

\end{abstract}



\section{Introduction}
\label{sec:introduction}

The angle between the orbital planes of planets, the mutual inclination, is an important proxy for studying the architecture of multi-planet systems. Mutual inclinations can offer valuable insights into planet formation and evolution history \citep[e.g.,][]{zhu_dong_review}. Theoretically, inclined orbits can arise through various mechanisms, such as gravitational scattering of planetary embryos \citep[][]{Hansen_Murray2013, Moriarty_Ballard2016, Izidoro2021} or from gravitational perturbations like distant giant planets \citep[][]{Mustill2017, Lai_Pu2017, Pu_Lai2018}. Besides planet-planet interactions, the stellar oblateness of a tilted star can induce differential nodal precession among coplanar planets, potentially exciting mutual inclinations after the protoplanetary disk has dissipated \citep{Millholland2021}. 

However, mutual inclinations are challenging to measure in practice for the majority of the exoplanet detection methods commonly used today \citep{zhu_dong_review}. In principle, one can infer inclinations using the transit method or astrometry, but the transit method is heavily biased toward coplanar systems, while astrometry is primarily sensitive to objects several AU away and requires an extended time baseline to achieve sufficient accuracy.  

Planets with short orbital periods provide a unique opportunity to investigate mutual inclinations, as their proximity to the host star results in a high chance of transit, even for highly inclined orbits. \cite{Dai2018} analyzed approximately 100 systems hosting short-period planets and concluded that short-period planets with $a/R_{\star}<5$ ($a$, orbital semi-major axis; $R_{\star}$, stellar radius) are more mutually inclined. This suggests that the processes responsible for shrinking these orbits also excite their inclinations, implying that these short-period planets have undergone dynamically chaotic, or ``hot'', evolution processes.

Recent studies indicate that dynamically hot systems preferentially reside around metal-rich stars \citep[e.g., see review by][]{pp7review2023}. Numerous works have demonstrated that giant planets are more common around metal-rich hosts \citep[e.g.,][]{Planet-Metallicity_Correlation, Johnson2010, Buchhave2012, Chen2023}, particularly those with eccentric orbits \citep{Dawson2013, Buchhave2018}. A similar preference for metal-rich stars also exists for small planets in occurrence rate \citep[][]{Wang2015, Zhu2019} and orbital eccentricity \citep{Mills2019, An2023}. \cite{An2023} found a positive correlation between metallicity and system-averaged mutual inclinations using the Transit Duration Ratio (TDR) method. In contrast, evidence suggests that compact multi-planet systems, which are generally more aligned \citep[e.g.,][]{CKS-V-Peas_in_a_pod}, tend to be around metal-poor stars \citep{Brewer2018, Anderson2021}. This implies that metal-poor systems might be dynamically cooler. 

In this paper, we report the correlation between mutual inclinations of the innermost two planets and stellar metallicities for short-period small planets ($a/R_{\star}<12$, $R<4\ \rm R_{\oplus}$), combining a sample of \textit{Kepler} \citep{Kepler_intro} and TESS (Transiting Exoplanet Survey Satellite, \citealt{TESS_Ricker2015}) discoveries and enabled by uniform catalogs of stellar metallicities. This paper is organized as follows: Section~\ref{sec:sample_selection} describes the sample selection process; Section~\ref{sec:stellar_pars} introduces our sources of stellar parameters; Section~\ref{sec:lc_fitting} details the photometric analyses to estimate mutual inclinations; and Section~\ref{sec:relation} reports our main findings and conclusions, followed by a discussion in Section~\ref{sec:discussion}.

\section{Sample Selection}
\label{sec:sample_selection}

We started with the sample from \cite{Dai2018}, in which they selected transiting short-period planets from \textit{Kepler} and K2 and conducted their own photometry analyses to determine planet properties. 
Specifically, \cite{Dai2018} required the apparent \textit{Kepler} magnitude $K_p < 14$~mag, and the innermost planet to have a radius smaller than 4~$\rm R_{\oplus}$, an $a/R_{\star} < 12$, and a transit signal-to-noise ratio $S/N>20$. The limit on $a/R_{\star} = 12$ corresponds to an allowed range of inclinations of $85-90^{\circ}$. We selected 94 systems with $a/R_{\star} < 12$ among the total 102 systems with innermost planets' $a/R_{\star} < 20$ in their sample. We further excluded two systems, KOI-1239 and EPIC 211305568, as these were classified as planet candidates which have not been confirmed according to the NASA Exoplanet Archive\footnote{We used the Planetary Systems Composite Data Table, available at: \url{https://exoplanetarchive.ipac.caltech.edu/cgi-bin/TblView/nph-tblView?app=ExoTbls&config=PSCompPars}, accessed on Mar. 8, 2024.} (hereafter, NEA), resulting in a final set of 92 systems.

We expanded the \cite{Dai2018} sample by adding TESS targets. We applied the same selection criteria to the confirmed TESS multi-transiting systems from NEA, which resulted in an addition of 18 systems. We did not apply magnitude or $S/N$ cuts to our TESS sample, as TESS targets are generally brighter than \textit{Kepler} stars \citep[e.g.][]{TOI_catalogpaper}, and they are all confirmed planets with comparable error bars in the measured inclinations to the \cite{Dai2018} sample. The faintest target in our sample, LP 791-18, has a TESS magnitude of 13.6, with the second faintest at 12.3 and all others brighter than 11. 

Finally, we restricted the size of the second innermost planets to be smaller than $4\ \rm R_{\oplus}$ to exclude giant planets in the inner system and focus on the mutual inclinations between small planets.\footnote{Beyond the innermost two planets, there are 6 systems in our sample known to have planets larger than $4\ \rm R_{\oplus}$ but all smaller than $6.2\ \rm R_{\oplus}$ (mostly between 4--5~$\rm R_{\oplus}$) except for Kepler-1131~d ($R\sim 9.7\ \rm R_{\oplus}$). Removing these systems does not affect our conclusions.} As a result, we selected a total of 89 systems for this study, consisting of 75 \textit{Kepler} and K2 systems and 14 TESS systems. We modeled the light curves of the TESS targets to derive the best-fit orbital parameters (Section~\ref{sec:lc_fitting}). For the TESS sample, the minimum value of $a/R_{\star}$ is about 2.55, which corresponds to a $\Delta i$ up to 21 degrees.

\begin{figure*}
    \centering
	\includegraphics[width=115mm]{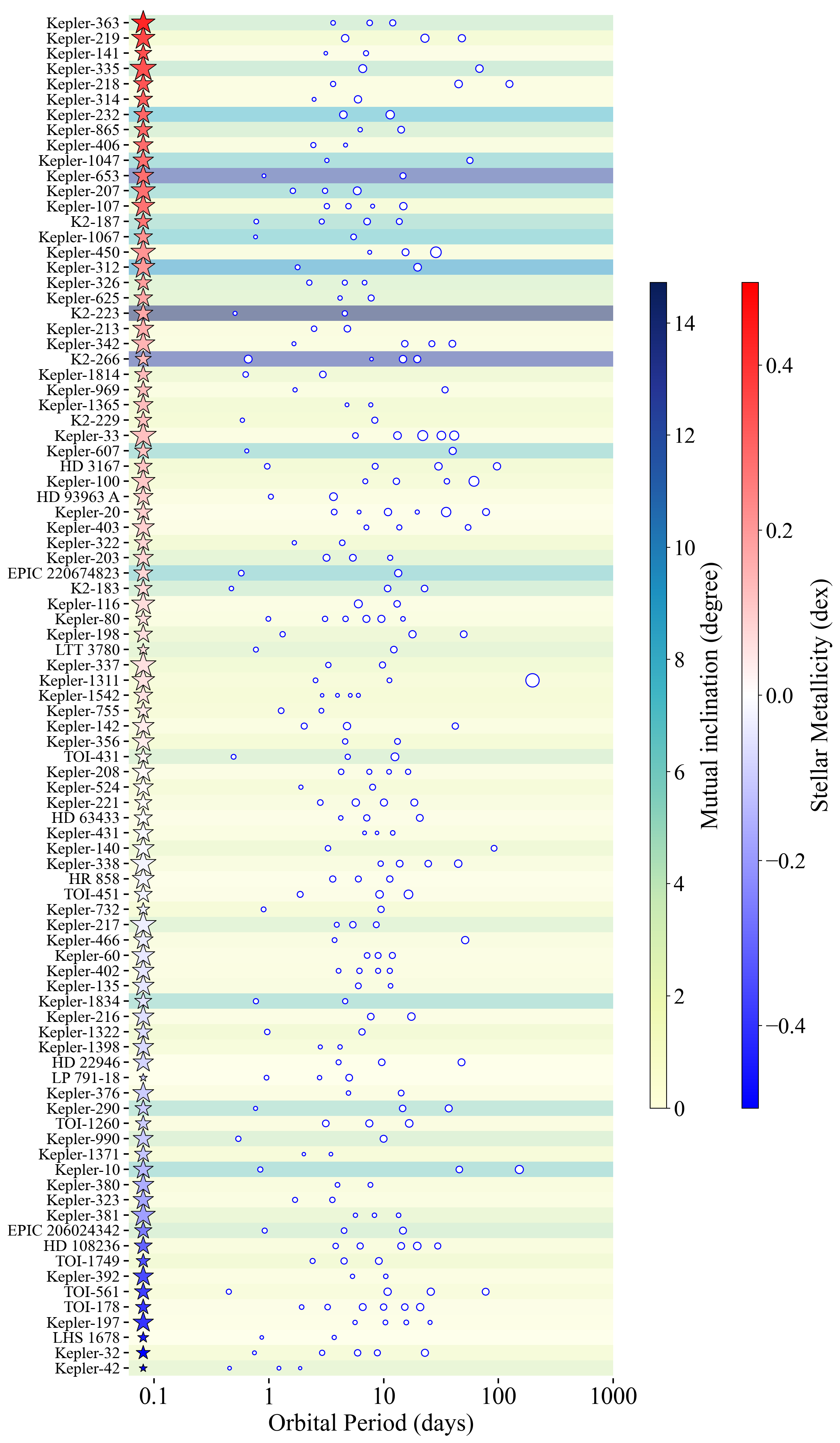}
    \caption{String plot for the 89 planetary systems in this work, with system names labeled on the left. Host stars are sorted and color-coded based on their [Fe/H] values, with more metal-rich stars plotted on top with warmer colors. The background grids are shaded according to mutual inclinations, $\Delta i$, where darker shades correspond to higher values. The sizes of the star and planet symbols are scaled by their radii. [Fe/H] and $\Delta i$ values used to create this figure are listed in Table~\ref{tab:systeminfo}, while other parameters are from the NEA. }
    \label{fig:stringplot}
\end{figure*}

For both the \cite{Dai2018} and TESS samples, we derived the mutual inclination ($\Delta i$) between the innermost two planets in each system using their best-fit orbital inclinations, assuming they have parallel transit chords going across the same stellar hemisphere, i.e. $\Delta i = |i_1 - i_2|$. This results in the lower limit of the mutual inclination between the two planets (Section~\ref{sec:discuss_caveats}). Table~\ref{tab:systeminfo} summarizes the information on these systems with an illustration in Figure~\ref{fig:stringplot}. 


\begin{longtable}{lccc}
\caption{Summary of the sample information} \label{tab:systeminfo} \\
\hline
\hline
Host  & [Fe/H] & $\Delta i$ & Origin of \\ 
 & (dex) & ($^{\circ}$) & [Fe/H] \\
\hline
\endfirsthead
\hline
Host & [Fe/H] & $\Delta i$ & Origin of \\ 
 & (dex) & ($^{\circ}$) & [Fe/H] \\ 
\hline
\endhead
\hline
\multicolumn{3}{l}{\textit{To be continued}} \\
\endfoot
\hline
\endlastfoot
EPIC 206024342 & ${-0.255}^{+0.011}_{-0.011}$ & ${3.95}^{+3.05}_{-3.06}$ & APOGEE \\
EPIC 220674823 & ${0.08}^{+0.006}_{-0.006}$ & ${6.31}^{+4.02}_{-4.82}$ & APOGEE \\
HD 3167 & ${0.114}^{+0.01}_{-0.01}$ & ${2.04}^{+1.46}_{-1.54}$ & APOGEE \\
K2-183 & ${0.078}^{+0.025}_{-0.025}$ & ${4.19}^{+3.93}_{-3.27}$ & LAMOST \\
K2-187 & ${0.273}^{+0.032}_{-0.032}$ & ${5.42}^{+1.78}_{-2.6}$ & LAMOST \\
K2-223 & ${0.17}^{+0.01}_{-0.01}$ & ${14.72}^{+1.5}_{-2.11}$ & SWEET-Cat \\
K2-229 & ${0.123}^{+0.025}_{-0.025}$ & ${1.84}^{+2.26}_{-1.28}$ & LAMOST \\
K2-266 & ${0.135}^{+0.029}_{-0.029}$ & ${12.67}^{+0.68}_{-0.75}$ & LAMOST \\
Kepler-10 & ${-0.139}^{+0.052}_{-0.052}$ & ${5.82}^{+0.16}_{-0.17}$ & PASTEL \\
Kepler-100 & ${0.107}^{+0.036}_{-0.036}$ & ${1.66}^{+0.46}_{-0.36}$ & LAMOST \\
Kepler-1047 & ${0.285}^{+0.07}_{-0.07}$ & ${6.3}^{+3.55}_{-2.84}$ & PASTEL \\
Kepler-1067 & ${0.213}^{+0.053}_{-0.053}$ & ${6.75}^{+1.67}_{-2.6}$ & PASTEL \\
Kepler-107 & ${0.275}^{+0.043}_{-0.043}$ & ${1.62}^{+0.8}_{-0.94}$ & PASTEL \\
Kepler-116 & ${0.07}^{+0.057}_{-0.057}$ & ${0.71}^{+1.71}_{-0.59}$ & PASTEL \\
Kepler-1311 & ${0.057}^{+0.053}_{-0.053}$ & ${1.53}^{+1.6}_{-1.03}$ & PASTEL \\
Kepler-1322 & ${-0.07}^{+0.151}_{-0.184}$ & ${2.1}^{+1.89}_{-1.5}$ & $\rm Morton2016^1$ \\
Kepler-135 & ${-0.05}^{+0.03}_{-0.03}$ & ${1.01}^{+0.93}_{-0.71}$ & PASTEL \\
Kepler-1365 & ${0.125}^{+0.03}_{-0.03}$ & ${2.09}^{+1.54}_{-0.95}$ & PASTEL \\
Kepler-1371 & ${-0.11}^{+0.03}_{-0.03}$ & ${1.59}^{+1.65}_{-1.08}$ & PASTEL \\
Kepler-1398 & ${-0.073}^{+0.083}_{-0.083}$ & ${1.45}^{+1.29}_{-0.98}$ & PASTEL \\
Kepler-140 & ${-0.015}^{+0.03}_{-0.03}$ & ${2.31}^{+0.57}_{-1.06}$ & PASTEL \\
Kepler-141 & ${0.345}^{+0.025}_{-0.025}$ & ${0.52}^{+0.53}_{-0.36}$ & PASTEL \\
Kepler-142 & ${0.037}^{+0.057}_{-0.057}$ & ${0.89}^{+0.81}_{-0.62}$ & PASTEL \\
Kepler-1542 & ${0.055}^{+0.043}_{-0.043}$ & ${1.77}^{+1.88}_{-1.34}$ & PASTEL \\
Kepler-1814 & ${0.13}^{+0.03}_{-0.03}$ & ${2.27}^{+1.46}_{-1.49}$ & PASTEL \\
Kepler-1834 & ${-0.057}^{+0.053}_{-0.053}$ & ${5.59}^{+1.3}_{-1.86}$ & PASTEL \\
Kepler-197 & ${-0.398}^{+0.043}_{-0.043}$ & ${0.35}^{+0.45}_{-0.24}$ & PASTEL \\
Kepler-198 & ${0.065}^{+0.06}_{-0.06}$ & ${2.43}^{+1.41}_{-1.91}$ & PASTEL \\
Kepler-20 & ${0.096}^{+0.058}_{-0.058}$ & ${0.2}^{+0.35}_{-0.16}$ & LAMOST \\
Kepler-203 & ${0.083}^{+0.09}_{-0.09}$ & ${3.39}^{+1.18}_{-1.3}$ & PASTEL \\
Kepler-207 & ${0.28}^{+0.023}_{-0.023}$ & ${5.93}^{+1.91}_{-2.33}$ & PASTEL \\
Kepler-208 & ${0.015}^{+0.025}_{-0.025}$ & ${0.68}^{+0.71}_{-0.43}$ & PASTEL \\
Kepler-213 & ${0.16}^{+0.053}_{-0.053}$ & ${0.94}^{+0.37}_{-0.37}$ & PASTEL \\
Kepler-216 & ${-0.06}^{+0.03}_{-0.03}$ & ${0.65}^{+0.59}_{-0.47}$ & PASTEL \\
Kepler-217 & ${-0.033}^{+0.023}_{-0.023}$ & ${3.53}^{+1.52}_{-1.86}$ & PASTEL \\
Kepler-218 & ${0.335}^{+0.055}_{-0.055}$ & ${0.8}^{+0.81}_{-0.58}$ & PASTEL \\
Kepler-219 & ${0.36}^{+0.043}_{-0.043}$ & ${1.85}^{+0.36}_{-0.43}$ & PASTEL \\
Kepler-221 & ${0.01}^{+0.03}_{-0.03}$ & ${0.98}^{+0.51}_{-0.51}$ & PASTEL \\
Kepler-232 & ${0.3}^{+0.1}_{-0.1}$ & ${7.55}^{+4.07}_{-5.61}$ & PASTEL \\
Kepler-290 & ${-0.1}^{+0.04}_{-0.04}$ & ${5.23}^{+2.37}_{-3.28}$ & PASTEL \\
Kepler-312 & ${0.2}^{+0.053}_{-0.053}$ & ${9.12}^{+1.19}_{-1.38}$ & PASTEL \\
Kepler-314 & ${0.32}^{+0.05}_{-0.05}$ & ${0.79}^{+0.73}_{-0.52}$ & PASTEL \\
Kepler-32 & ${-0.5}^{+0.32}_{-0.32}$ & ${1.64}^{+1.33}_{-1.16}$ & PASTEL \\
Kepler-322 & ${0.083}^{+0.053}_{-0.053}$ & ${2.12}^{+0.92}_{-1.21}$ & PASTEL \\
Kepler-323 & ${-0.173}^{+0.053}_{-0.053}$ & ${0.82}^{+1.01}_{-0.55}$ & PASTEL \\
Kepler-326 & ${0.19}^{+0.025}_{-0.025}$ & ${3.59}^{+1.36}_{-1.51}$ & PASTEL \\
Kepler-33 & ${0.122}^{+0.066}_{-0.066}$ & ${0.76}^{+1.02}_{-0.53}$ & PASTEL \\
Kepler-335 & ${0.34}^{+0.07}_{-0.07}$ & ${4.54}^{+0.34}_{-0.31}$ & PASTEL \\
Kepler-337 & ${0.06}^{+0.053}_{-0.053}$ & ${2.16}^{+1.27}_{-1.31}$ & PASTEL \\
Kepler-338 & ${-0.02}^{+0.056}_{-0.056}$ & ${1.0}^{+0.58}_{-0.66}$ & PASTEL \\
Kepler-342 & ${0.147}^{+0.043}_{-0.043}$ & ${0.85}^{+1.19}_{-0.58}$ & PASTEL \\
Kepler-356 & ${0.037}^{+0.053}_{-0.053}$ & ${1.8}^{+0.3}_{-0.32}$ & PASTEL \\
Kepler-363 & ${0.422}^{+0.043}_{-0.043}$ & ${4.09}^{+1.36}_{-1.06}$ & PASTEL \\
Kepler-376 & ${-0.1}^{+0.053}_{-0.053}$ & ${0.8}^{+0.97}_{-0.6}$ & PASTEL \\
Kepler-380 & ${-0.16}^{+0.103}_{-0.103}$ & ${1.1}^{+1.54}_{-0.78}$ & PASTEL \\
Kepler-381 & ${-0.187}^{+0.053}_{-0.053}$ & ${2.77}^{+1.44}_{-1.81}$ & PASTEL \\
Kepler-392 & ${-0.355}^{+0.03}_{-0.03}$ & ${0.78}^{+0.77}_{-0.52}$ & PASTEL \\
Kepler-402 & ${-0.05}^{+0.023}_{-0.023}$ & ${0.77}^{+0.74}_{-0.5}$ & PASTEL \\
Kepler-403 & ${0.093}^{+0.023}_{-0.023}$ & ${0.56}^{+0.63}_{-0.4}$ & PASTEL \\
Kepler-406 & ${0.289}^{+0.025}_{-0.025}$ & ${1.02}^{+0.97}_{-0.77}$ & LAMOST \\
Kepler-42 & ${-0.5}^{+0.09}_{-0.09}$ & ${2.95}^{+0.6}_{-0.77}$ & $\rm Mann2017^2$ \\
Kepler-431 & ${-0.013}^{+0.043}_{-0.043}$ & ${0.9}^{+1.13}_{-0.65}$ & PASTEL \\
Kepler-450 & ${0.21}^{+0.065}_{-0.065}$ & ${0.99}^{+0.38}_{-0.4}$ & PASTEL \\
Kepler-466 & ${-0.037}^{+0.053}_{-0.053}$ & ${1.02}^{+0.81}_{-0.71}$ & PASTEL \\
Kepler-524 & ${0.01}^{+0.03}_{-0.03}$ & ${1.65}^{+1.82}_{-1.16}$ & PASTEL \\
Kepler-60 & ${-0.047}^{+0.023}_{-0.023}$ & ${0.73}^{+1.08}_{-0.53}$ & PASTEL \\
Kepler-607 & ${0.12}^{+0.04}_{-0.04}$ & ${5.89}^{+3.06}_{-3.96}$ & PASTEL \\
Kepler-625 & ${0.173}^{+0.053}_{-0.053}$ & ${3.34}^{+1.08}_{-1.25}$ & PASTEL \\
Kepler-653 & ${0.283}^{+0.053}_{-0.053}$ & ${12.38}^{+1.52}_{-1.72}$ & PASTEL \\
Kepler-732 & ${-0.029}^{+0.008}_{-0.008}$ & ${1.71}^{+0.55}_{-0.67}$ & APOGEE \\
Kepler-755 & ${0.04}^{+0.07}_{-0.07}$ & ${1.58}^{+2.67}_{-1.07}$ & PASTEL \\
Kepler-80 & ${0.065}^{+0.055}_{-0.055}$ & ${0.89}^{+0.9}_{-0.65}$ & PASTEL \\
Kepler-865 & ${0.3}^{+0.12}_{-0.12}$ & ${3.91}^{+2.33}_{-2.69}$ & PASTEL \\
Kepler-969 & ${0.13}^{+0.03}_{-0.03}$ & ${0.82}^{+0.96}_{-0.6}$ & PASTEL \\
Kepler-990 & ${-0.103}^{+0.053}_{-0.053}$ & ${3.82}^{+3.89}_{-2.96}$ & PASTEL \\
HD 108236$^*$ & ${-0.302}^{+0.011}_{-0.011}$ & ${1.18}^{+0.67}_{-0.48}$ & SWEET-Cat \\
HD 22946$^*$ & ${-0.085}^{+0.015}_{-0.015}$ & ${0.12}^{+0.97}_{-0.89}$ & SWEET-Cat \\
HD 63433$^*$ & ${0.0}^{+0.02}_{-0.02}$ & ${0.41}^{+0.8}_{-0.88}$ & SWEET-Cat \\
HD 93963 A$^*$ & ${0.1}^{+0.04}_{-0.04}$ & ${0.77}^{+0.99}_{-1.27}$ & SWEET-Cat \\
HR 858$^*$ & ${-0.02}^{+0.02}_{-0.02}$ & ${0.24}^{+0.5}_{-0.62}$ & SWEET-Cat \\
LHS 1678$^*$ & ${-0.45}^{+0.11}_{-0.11}$ & ${0.11}^{+0.8}_{-0.95}$ & SWEET-Cat \\
LP 791-18$^*$ & ${-0.09}^{+0.19}_{-0.19}$ & ${0.08}^{+0.91}_{-0.89}$ & $\rm Peterson2023^3$ \\
LTT 3780$^*$ & ${0.06}^{+0.11}_{-0.11}$ & ${3.27}^{+0.63}_{-0.56}$ & $\rm Bonfanti2024^4$ \\
TOI-1260$^*$ & ${-0.1}^{+0.07}_{-0.07}$ & ${0.92}^{+0.46}_{-0.36}$ & $\rm Lam2023^5$ \\
TOI-1749$^*$ & ${-0.347}^{+0.014}_{-0.014}$ & ${2.11}^{+0.54}_{-0.46}$ & APOGEE \\
TOI-178$^*$ & ${-0.393}^{+0.053}_{-0.053}$ & ${0.43}^{+0.97}_{-0.94}$ & SWEET-Cat \\
TOI-431$^*$ & ${0.02}^{+0.08}_{-0.08}$ & ${3.85}^{+1.51}_{-1.48}$ & PASTEL \\
TOI-451$^*$ & ${-0.025}^{+0.019}_{-0.019}$ & ${0.41}^{+0.99}_{-1.09}$ & SWEET-Cat \\
TOI-561$^*$ & ${-0.378}^{+0.008}_{-0.008}$ & ${1.34}^{+1.32}_{-1.64}$ & APOGEE \\
\end{longtable}
\tablenotetext{*}{Mutual inclinations measured from this work.} 
\tablenotetext{1}{\cite{Morton2016}} 
\tablenotetext{2}{\cite{Mann2017}} \tablenotetext{3}{\cite{Peterson2023}}
\tablenotetext{4}{\cite{Bonfanti2024}}
\tablenotetext{5}{\cite{Lam2023}}

\section{Stellar Parameters}
\label{sec:stellar_pars}

We collected the stellar properties for our targets by cross-matching the samples with the Planetary Systems Composite Data table from NEA, in particular, the metallicity, [Fe/H], and mass of each host star. However, these measurements come from various publications and are thus heterogeneous. In order to minimize biases that may arise from different methods of estimating [Fe/H], we searched for [Fe/H] values for our targets across various databases of stellar parameters, including LAMOST (The Large Sky Area Multi-Object Fiber Spectroscopic Telescope, \citealt{lamost}), APOGEE (Apache Point Observatory Galactic Evolution Experiment, \citealt{APOGEE}), SWEET-Cat \citep{sweet-cat2021} and PASTEL \citep{pastel2016}. 

Briefly, PASTEL compiles atmospheric parameters (effective temperature $T_{\rm eff}$, surface gravity log~$g$, and [Fe/H]) from high-resolution, high-S/N spectra for over 18,000 stars with [Fe/H] precision better than 0.05 dex for stars with [Fe/H] > $-2.0$, as noted by \cite{huangyang2022}. SWEET-Cat provides homogeneously derived spectroscopic parameters and utilizes Gaia eDR3 parallaxes to improve surface gravity precision, thus enhancing the reliability of the catalog’s stellar parameters. The LAMOST and APOGEE catalogs are commonly used for exoplanet demographic studies as their parameters are uniformly derived using data from large spectroscopic surveys \citep[e.g.,][]{Xie2016}. 

\begin{figure*}
    \centering
	\includegraphics[width=115mm]{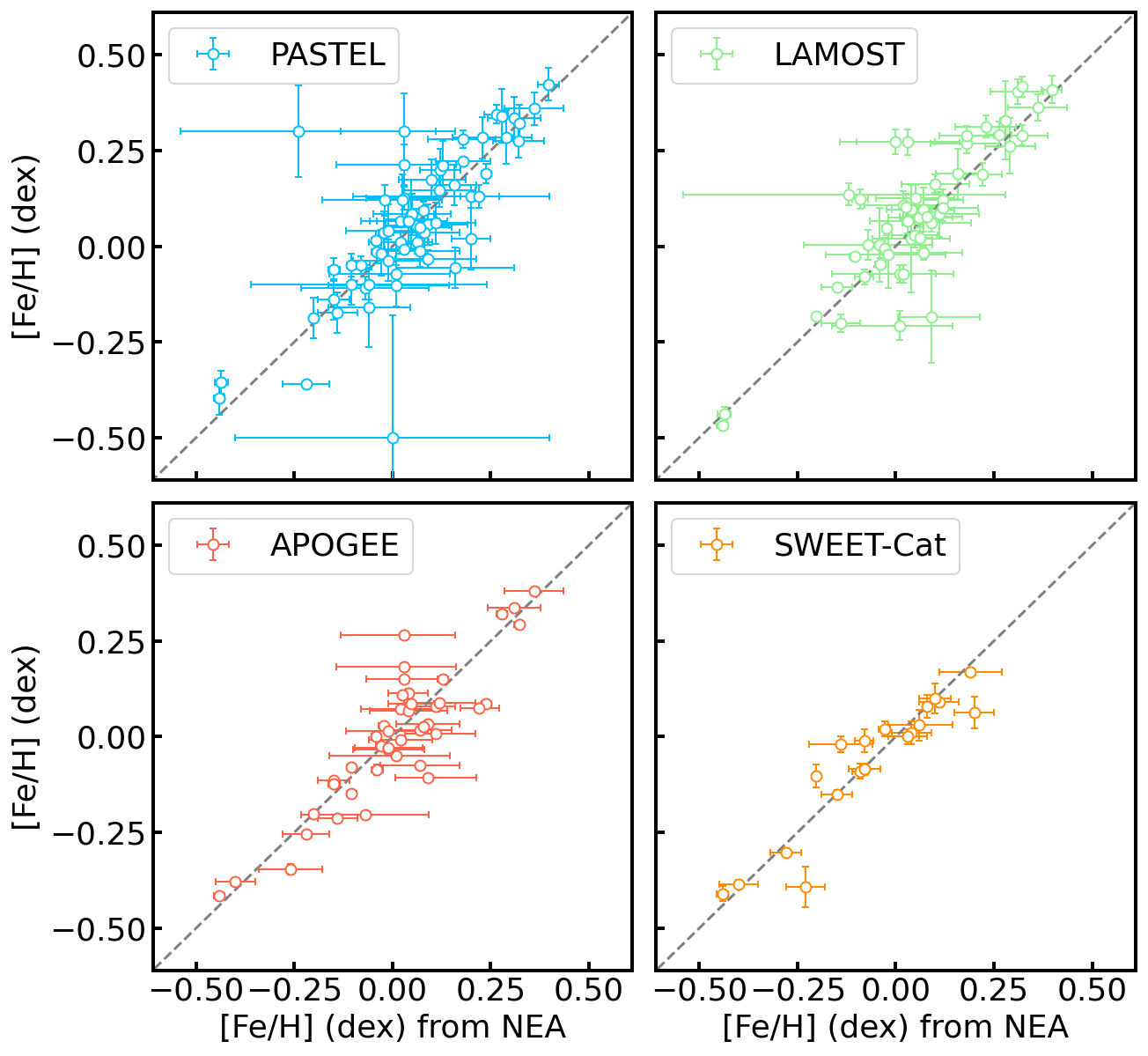}
    \caption{Comparison of stellar metallicity ([Fe/H]) between each database (y-axis) and the literature (x-axis) for overlapping systems in our sample. The measurements show good agreement with published data within the error bars, with no obvious offset or bias recognized. }
    \label{fig:feh_compare}
\end{figure*}

We cross-matched our sample with these databases and found 67 matches in PASTEL, 52 in LAMOST, 43 in APOGEE, and 21 in SWEET-Cat. Figure~\ref{fig:feh_compare} compares the [Fe/H] values of the cross matches from each database with the literature values listed on NEA. Among the databases, PASTEL showed no significant bias or offset compared to the uniform databases (LAMOST and APOGEE) and covered the most targets in our sample. Therefore, we prioritized using the [Fe/H] from these databases as follows: PASTEL $>$ LAMOST $>$ APOGEE $>$ SWEET-Cat $>$ Literature, favoring results from high-resolution spectroscopy (PASTEL). We also verified that our conclusion remained if we primarily adopted the metallicities from the uniform catalogs (LAMOST and APOGEE).


In addition, we have five targets using [Fe/H] values from the literature as they are not covered by the databases listed above. We have also verified that our main conclusion (next section) does not depend on the specific choice or prioritization of [Fe/H] values, as long as they are mostly from these databases. We list the metallicities of the 89 targets in our sample and their origins in Table~\ref{tab:systeminfo}.

\section{Mutual inclinations from TESS photometry}
\label{sec:lc_fitting}

We performed fitting to the light curves of the 14 TESS systems to ensure a homogeneous analysis in deriving $\Delta i$, similar to \cite{Dai2018}, and we describe our fittings in this section. We downloaded all available light curves with 2-min exposures for the TESS targets reduced by the Science Processing Operations Center \citep[SPOC;][]{Jenkins2016} from the Mikulski Archive for Space Telescopes\footnote{\url{https://archive.stsci.edu}} (MAST) using the python package \code{lightkurve} \citep{lightkurve}. We used the Presearch Data Conditioning Simple Aperture Photometry light curve files \citep[PDCSAP flux;][]{Stumpe2012, Smith2012, Stumpe2014}, which were extracted using Simple Aperture Photometry (SAP) and calibrated for instrument systematics by the Presearch Data Conditioning (PDC) algorithm. 



We first detrended the light curves using the Gaussian Process (GP) models with the Matern 3/2 kernel, and all transit windows were masked according to the transit epochs and periods from the literature. We set uninformative priors for the GP with wide bounds for most parameters. We examined the detrended light curves and identified two systems around active stars, HD 63433 and TOI-451. We further processed their light curves to eliminate the stellar flares by removing data points beyond the 2.5--3~$\sigma$ range of a smoothed version of the light curves before performing another round of detrending. 




After detrending, we performed an initial transit fit using the \code{Juliet} package \citep{Juliet}, accounting for all known transiting planets in each system. We assume Gaussian priors for the transit-time-center, $t_{0,i}$, and period, $P_i$, according to the best-fit values reported in the literature, with the standard deviation set to 1.5 times the transit duration. To simplify the initial model, we fixed eccentricities, $e_i$, and arguments of periapsis, $\omega_i$, for all planets to zero and 90~degrees, respectively. The host star density, $\rho_{\star}$, was assigned a normal distribution prior using the reported value and uncertainties from publications. All other stellar, planetary, or light curve systematic parameters were set to have uninformative priors (wide uniform or log-uniform distributions).


If the best-fit planetary parameter has a relatively large discrepancy  (e.g. $> 1\sigma$) with the literature, we would adopt more informative priors for a second round of fitting in order to better constrain the orbital parameters. The specific adjustments on the priors vary from target to target. For instance, we would narrow the priors of $t_{0,i}$ or period or planet-to-star radius ratio, or adopt the exact same priors as in the publications. 
We examined the best-fit phase-folded light curves to confirm that none of our systems has detectable TTVs.

Finally, we performed a model comparison between circular and eccentric orbit configurations. We used the parameterization of $\sqrt{e_i}\sin \omega_i$ and $\sqrt{e_i}\cos \omega_i$ with priors $\mathcal{U}(-0.6, 0.6)$ (a somewhat arbitrary value smaller than 1.0 for computational efficiency), corresponding to eccentricities ranging from 0 to 0.72. Only one system, TOI-451, showed a strong preference for the eccentric orbit model, with $\Delta(\log Z) = 51.5$. We then adopted the eccentric orbit model, though the mutual inclination remains consistent with the circular one within the uncertainty. The $\Delta i$ values derived from the best-fit inclinations via $\Delta i = |i_1 - i_2|$ are listed in Table~\ref{tab:systeminfo}.

\section{Mutual Inclination vs. Metallicity}
\label{sec:relation}

\subsection{Main results}
\label{sec:relation_main}

The stellar metallicity of the host star and the mutual inclinations, $\Delta i$, for the innermost two planets in each of the 89 systems in our sample are listed in Table~\ref{tab:systeminfo}. We emphasize that this $\Delta i$ that calculated from $|i_1 - i_2|$ is the lower limit of the true mutual inclination (see discussion in Section~\ref{sec:discuss_caveats}). 

\begin{figure*}
    \centering
	\includegraphics[width=120mm]{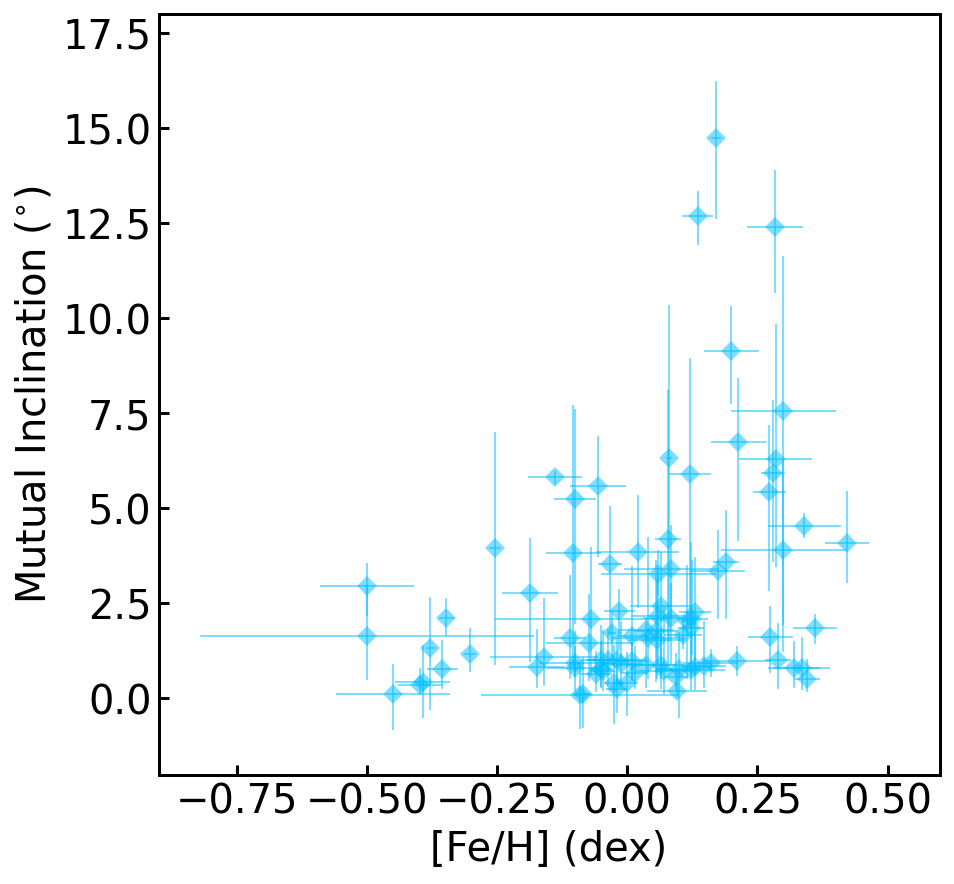}
    \caption{Stellar metallicity vs. mutual inclination plot. The mutual inclinations come from light curve fitting, and [Fe/H] are carefully picked from multiple uniform databases (see Section~\ref{sec:stellar_pars}). In total, we have 89 systems in our sample. We found that as stellar metallicity increases, the distribution of mutual inclination tends to be more spread. }
    \label{fig:feh_deli}
\end{figure*}

The relation between $\Delta i$ and [Fe/H] is shown in Figure~\ref{fig:feh_deli}: systems with higher stellar metallicity tend to exhibit larger mutual inclinations. A Spearman correlation test yields a coefficient of 0.31 and a $p$-value of 0.0031, suggesting a moderate positive correlation between [Fe/H] and $\Delta i$. However, a more prominent feature of this trend is that the distribution of mutual inclinations is more dispersed for metal-rich hosts. For a quantitative comparison, we divided the sample into metal-rich and metal-poor groups based on the median [Fe/H] value of 0.055~dex, resulting in 45 metal-rich and 44 metal-poor systems. An Anderson-Darling (AD) test between these two groups yields a $p$-value of 0.0025, which indicates that they originate from distinct populations. 

\begin{figure*}
    \centering
	\includegraphics[width=170mm]{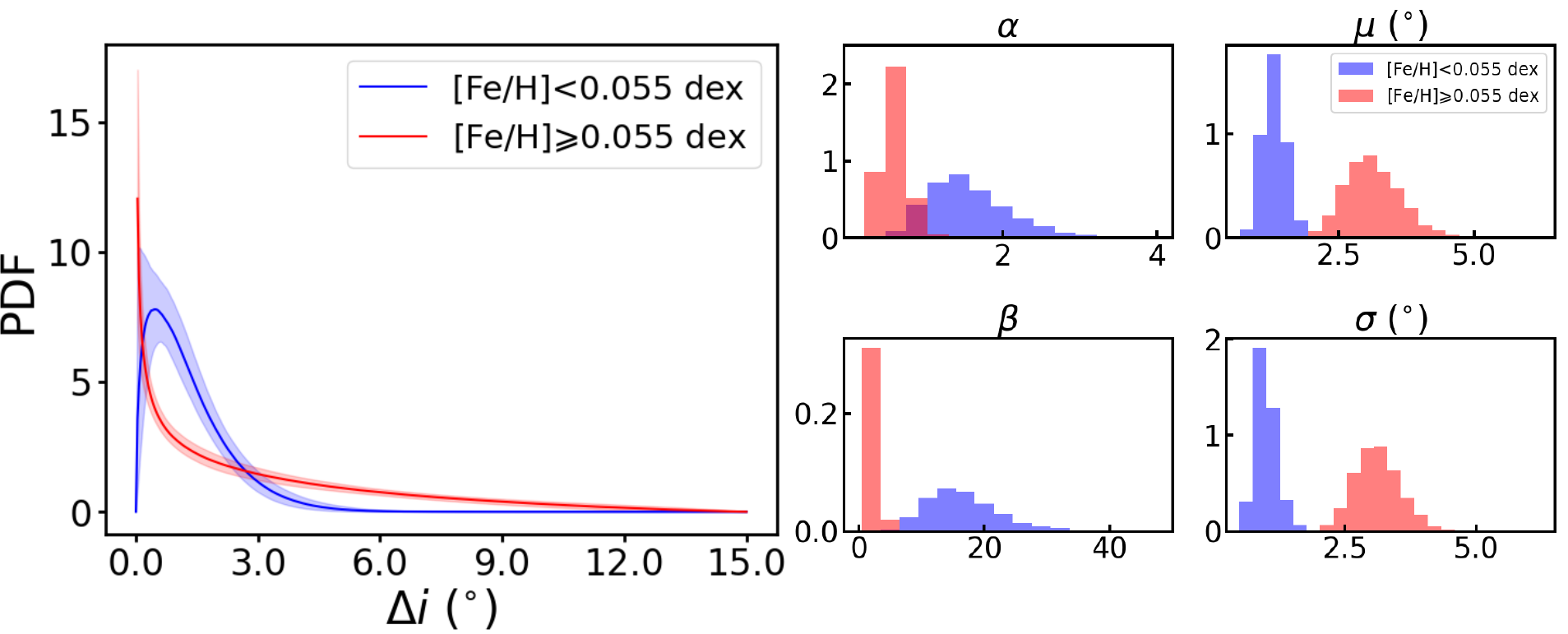}
    \caption{Distributions of $\Delta i$ of metal-poor (blue) and metal-rich (red) samples modeled by $\beta$-distributions. The posterior distributions of the model parameters are attached on the right, where $\mu$ is the mean value of the $\beta$-distribution derived using $\alpha$ and $\beta$, and $\sigma$ is the square root of the variance. The posterior distributions of the mean and variance reveal clear differences between the metal-poor and metal-rich samples. The metal-rich group exhibits a larger average $\Delta i$ and a broader $\Delta i$ distribution, with $\mu=3.13^{\circ}$$^{+0.51}_{-0.49}$ and $\sigma=3.12^{\circ}$$^{+0.43}_{-0.43}$, compared to the metal-poor group, which has $\mu=1.30^{\circ}$$^{+0.20}_{-0.21}$ and $\sigma=1.00^{\circ}$$^{+0.19}_{-0.19}$. }
    \label{fig:beta_distribution}
\end{figure*}

To further quantify the differences in the mutual inclination distributions between the metal-rich and metal-poor groups, we modeled the distribution of $\Delta i$ using a $\beta$-distribution. We adopted $\beta$-distributions in our model because of its flexibility in shape (e.g., in comparison to the Rayleigh distribution; \citealt{Lissauer2011}). We show the best-fit $\beta$-distribution models in Figure~\ref{fig:beta_distribution}, along with the posterior distributions of parameters, $\alpha$, and $\beta$, and the derived mean ($\mu$) and variance ($\sigma$) of $\Delta i$. The posteriors and Bayesian evidence were derived using the nested sampling Monte Carlo algorithm MLFriends \citep{Buchner2016, Buchner2019}, implemented through the \code{UltraNest}\footnote{\url{https://johannesbuchner.github.io/UltraNest/}} package \citep{Buchner2021}. The algorithm considers heterogeneous and asymmetric uncertainties for each data point.
As illustrated in Figure~\ref{fig:beta_distribution}, the metal-rich (red) and metal-poor (blue) groups exhibit remarkably different $\Delta i$ distributions. The metal-rich group shows a larger average $\Delta i$ with $\mu = 3.13^{\circ}$$^{+0.51}_{-0.49}$, and a more spread distribution characterized by $\sigma = 3.12^{\circ}$$^{+0.43}_{-0.43}$. In contrast, the metal-poor group has $\mu = 1.30^{\circ}$$^{+0.20}_{-0.21}$ and $\sigma = 1.00^{\circ}$$^{+0.19}_{-0.19}$, both of which are significantly smaller. 
The Mann-Whitney U test \citep{Mann_WhitneyUtest1947} gives $p$-values $<10^{-10}$ for both $\mu$'s and $\sigma$'s posterior distributions, indicating that the means of the two distributions are significantly different. 

\subsection{Other effects}
\label{sec:relation_others}

\begin{figure*}
    \centering
	\includegraphics[width=160mm]{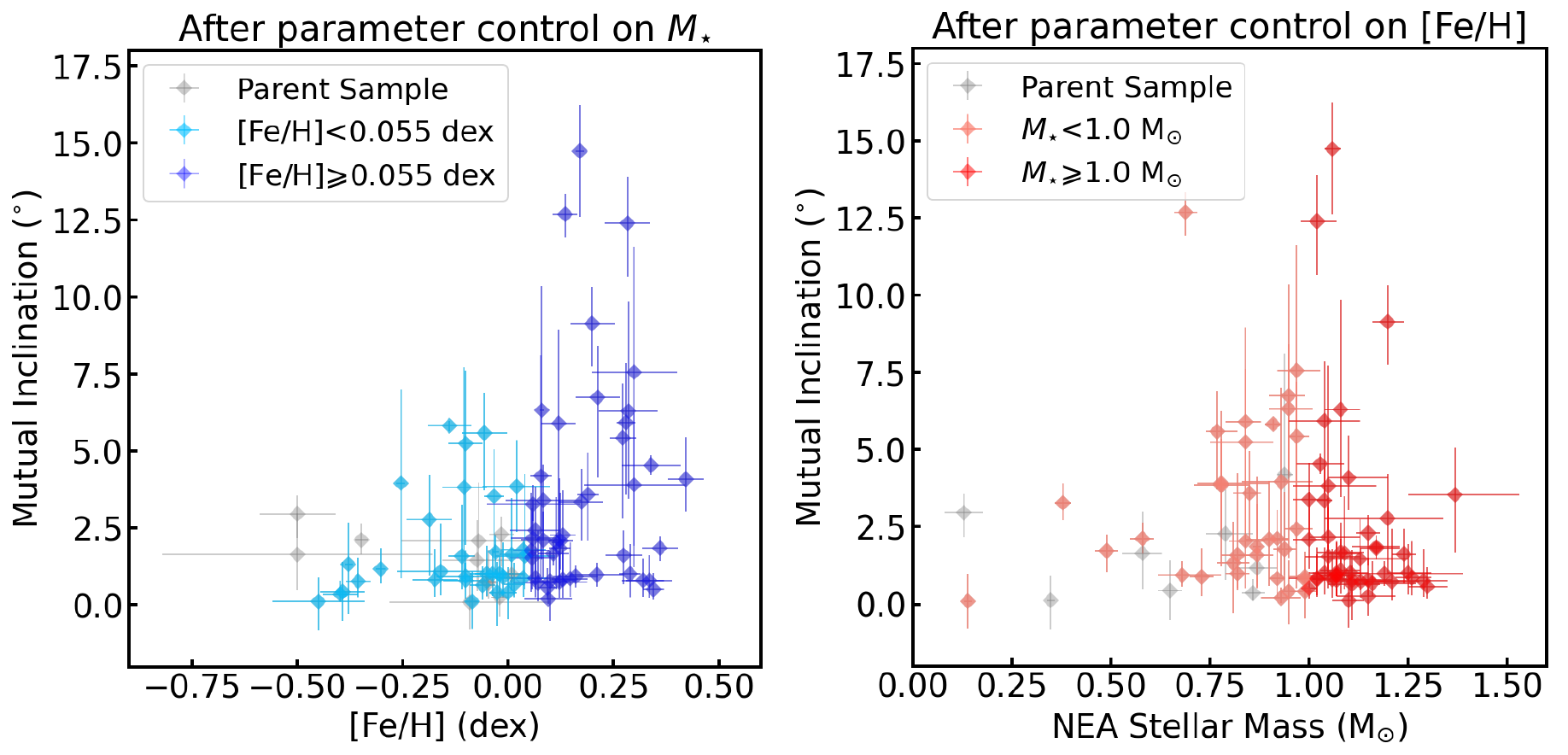}
    \caption{Relations between mutual inclination and stellar metallicity and stellar mass after sample control. Left: stellar metallicity vs. mutual inclination plot for a subsample of 78 systems selected to minimize the influence of stellar mass. Right: stellar mass vs. mutual inclination plot for a subsample of 81 systems selected to minimize the influence of stellar metallicity. Unlike [Fe/H] vs.\ $\Delta i$, the correlation between stellar mass and $\Delta i$ is not statistically significant in our sample, although we cannot completely rule out such a correlation. See Section~\ref{sec:relation} for more details.}
    \label{fig:par_ctrl}
\end{figure*}

Stellar metallicity is related to stellar mass. We considered the correlation between stellar mass and $\Delta i$ (right panel of Figure~\ref{fig:par_ctrl}) and how this could affect the [Fe/H]-$\Delta i$ correlation. The stellar mass measurements were adopted directly from the NEA. At first glance, higher mass stars seem to host planets with higher $\Delta i$. However, the Spearman correlation test yields $r = -0.17$ and $p = 0.11$, and the AD test between the low- and high-mass group divided by the median mass (1.0~$\rm M_{\odot}$, 44 in the sub-solar bin and 45 in the super-solar bin) of our sample gives $p = 0.15$, indicating low statistical significance. In addition, we performed the same $\beta$-distribution modeling to four sub-samples divided by the median [Fe/H] and stellar mass, and we found that $\Delta i$ varies more significantly along the metallicity dimension than the stellar mass one. 



To further verify the effect from stellar mass, we employed the nearest neighbor method for a sample control test to decouple the correlation between $\Delta i$ vs. stellar mass or metallicity, following a procedure similar to that of \cite{An2023}. Specifically, we first selected a star from the metal-rich group and calculated its mass difference with each star in the metal-poor group. We then matched the metal-rich star with the two metal-poor stars with the smallest mass differences. This process was repeated for every star in the metal-rich group, resulting in a new subsample from the metal-poor group with a mass distribution closely resembling that of the metal-rich group. This approach allowed us to minimize the influence of stellar mass on the relationship between $\Delta i$ and stellar metallicity. As a result, 33 stars were selected from the metal-poor group, creating a total subsample of 78 systems (left panel of Figure~\ref{fig:par_ctrl}).

A similar approach was applied to the subsample divided by stellar mass at the median value of 1.0~$\rm M_{\odot}$ to reduce the influence of metallicity (right panel of Figure~\ref{fig:par_ctrl}). We selected 35 systems from the sub-solar mass group, resulting in a new subsample of 81 systems. 
The same tests were conducted on the new samples. The Spearman test indicated that the correlation between $\Delta i$ and metallicity remains significant, with a $p$-value of 0.0028, and the AD test comparing the metal-rich and metal-poor subgroups yields a $p$-value of 0.0072. In contrast, for stellar mass, the Spearman and AD tests returned $p$-values of 0.02 and 0.089, respectively, indicating no significant evidence for a relationship between stellar mass and $\Delta i$. We conclude that the [Fe/H]-$\Delta i$ correlation is the dominant one, although our sample cannot fully rule out an additional correlation between $\Delta i$ and stellar mass. We also performed a similar parameter control on the stellar radius and arrived at the same conclusion.

Besides stellar mass, stellar metallicity is negatively correlated with stellar age \citep{Bensby2014, Chen2021}. However, the observed positive correlation between $\Delta i$ and [Fe/H] is unlikely to be due to the metallicity-age correlation. Both theories \citep{Zhou2007} and statistical studies \citep{Yang2023} have shown that mutual inclinations increase with stellar age, which would predict a negative correlation between $\Delta i$ and [Fe/H], contrary to what we observe here. Due to the limited number of samples with precise age measurements, parameter control cannot yet be effectively performed. Nevertheless, we expect that after correcting for the age effect, the observed positive correlation between the metallicity and mutual inclination will become stronger. 


Finally, we address the selection bias in metal-rich systems, as the metal-rich stars have larger radii in general than the metal-poor stars in our sample. This results in higher transit probabilities for multi-planet systems and thus larger detectable $\Delta i$ for the metal-rich stars. We calculated the maximum allowed mutual inclination, $\Delta i_{max}$\footnote{$\Delta i_{max}=\mathrm{arcsin}(R_{\star}/a_1)+\mathrm{arcsin}(R_{\star}/a_2)$, where $a_1$ and $a_2$ are the semi-major axes of the innermost and second innermost planets.}, for each system. The Spearman test yields $p=0.07$ between [Fe/H] and $\Delta i_{max}$, indicating a weak correlation (selection bias). We then performed a similar sample control on $\Delta i_{max}$ to further reduce the selection bias on our sample, in which 82 systems were reselected. The [Fe/H]-$\Delta i$ correlation remains significant with Spearman $p=0.0061$ and AD test $p=0.0079$. Best-fit $\beta$-distributions for the metal-rich ([Fe/H] $\geqslant0.0625$~dex) and metal-poor bins are $\mu=3.2^{\circ}\pm0.6$, $\sigma=3.2^{\circ}\pm0.5$ and $\mu=1.5^{\circ}\pm0.2$, $\sigma=1.1^{\circ}\pm0.2$, respectively, consistent with the results from the original sample. This indicates that this selection bias does not significantly affect our conclusions.

\section{Discussion}
\label{sec:discussion}


\subsection{Comparison with previous studies}
\label{sec:discuss_observations}

\cite{An2023} identified a positive mutual inclination-metallicity trend after examining a sample of small \textit{Kepler} planets/candidates with uniform [Fe/H] measurements from LAMOST. While our conclusion generally aligns with the finding in \cite{An2023}, it is important to note that our study focuses on a different sample from theirs: our innermost planets all have $a/R_{\star} < 12$ and \cite{An2023} used a generic \textit{Kepler} sample, with only 14\% among 152 systems having the innermost planet with $a/R_{\star} < 12$. Furthermore, they calculated average mutual inclinations, $\overline{\Delta i}$, for entire planetary systems across metallicity bins in a probabilistic manner, whereas we derived $\Delta i$ specifically between the innermost two planets. They used the mutual TDR as a proxy for mutual inclinations allowing more transit scenarios, which would lead to higher estimations on $\overline{\Delta i}$ than ours (see more in Section~\ref{sec:discuss_caveats}). However, Figure~13 in \cite{An2023} shows comparable mutual inclinations to ours, which means they would have derived lower values for $\Delta i$ if adopting the same assumption as ours, i.e. the planets transit over the same stellar hemisphere (Section~\ref{sec:sample_selection}). Unfortunately, the uncertainties on the sample means are relatively large for both samples (both being $\sim 3^{\circ} \pm (1\sim2)^{\circ}$), making it hard to conclude quantitatively. Nonetheless, the general consistency between the two works lends confidence to the robustness of our results, although it is not clear yet whether the mutual inclination-metallicity trend in the short-period planet sample is an extension of the generic trend or a distinct one. 



It is important to consider the metallicity-inclination correlation in tandem with orbital periods. \cite{Dai2018} observed that shorter-period planets tend to have higher mutual inclinations, while \cite{CKS-IV2018} and \cite{Mulders2016} reported that short-period planets ($P < 10$ days) are more common around metal-rich stars, which seems to suggest that the metallicity-inclination correlation reported here might be a projection of these two previously known correlations. We performed a test to show that this is not the case using a control sub-sample with orbital periods between 1--5 days, where the period-metallicity anti-correlation plateaus with the Spearman test yields $r=-0.05$ and $p=0.73$, indicating no significant correlation between periods and metallicities. For the 48 samples within this period range, the Spearman test yields $r=0.32$, $p=0.025$ between [Fe/H] and $\Delta i$, and the AD test yields $p=0.016$ between metal-rich and metal-poor bins divided by the mean [Fe/H] (0.056~dex) of the sub-sample, which indicate a consistent positive correlation, suggesting that the [Fe/H]-$\Delta i$ correlation should be intrinsic rather than a projection.

In addition, we highlight that the mutual inclination-metallicity correlation is not driven by the presence of USPs in our sample for the following reasons: (1) There is no evidence suggesting that USPs are associated with more metal-rich stars compared to other short-period planets with $P=1$--10~days \citep{Winn2017}. In our sample, the mean metallicity of the USP hosts does not appear to be super-solar. (2) The systems with USPs as the innermost planets do have higher mutual inclinations in general, but the relation persists after removing all USPs ($P < 1$~day) from our sample, for which the Spearman correlation coefficient $r=0.36$ with a p-value of 0.0042.

\subsection{Possible theoretical explanations}
\label{sec:discuss_theories}

The observed correlation among period, metallicity, and mutual inclination appears to favor a dynamically hot origin, similar to the formation of hot Jupiters. 

One possibility is that the disks of metal-rich stars could harbor more solids, leading to the formation of more giant planets \citep[e.g.,][]{Ida2008, Mordasini2012, Ndugu2018}. 
A mutually inclined distant giant can tilt the inner planets to inclined orbits via precession \citep[e.g.][]{Becker2017, Lai_Pu2017, Pu_Lai2018, Pu_Lai2021} or alternatively through planet-planet scattering \citep[e.g.][]{Mustill2017}. This picture seems to corroborate with the observational findings that super-Earths are often accompanied by outer giant planets \citep{zhu2018, Bryan2019}, especially for metal-rich stars \citep{Zhu2024}, but unfortunately, our sample does not have sufficient follow-up data to enable further investigation on the inner-outer correlation. 

Without invoking the existence of giant planets, high metallicity could result in a higher pebble flux in the protoplanetary disk, raising the probability of forming resonant chains \citep{pp7review2023}. This will end up in dynamical instability once the gas disk disperses, leading to a more dynamically hot system with higher $\Delta i$ and $e$ in general \citep[e.g., ][]{Izidoro2017, Lambrechts2019, Izidoro2021}.

Alternatively, secular chaos may produce highly inclined innermost planets in mature planetary systems (and could also explain USP formation, as in \citealt{Petrovich2019, Pu_Lai2019}). Secular interactions can redistribute angular momentum deficit (AMD) and systems with more and higher-mass planets typically have greater AMD. Considering that metal-rich systems tend to host more diverse planets in terms of size and mass \citep{Millholland2017, CKS-IV2018}, this may explain why such dynamical excitation is correlated with stellar metallicity, especially for the innermost planets. However, forming inclined USPs through secular chaos typically requires widely spaced external planets \citep[e.g.][]{Petrovich2019, Becker2020}, a configuration not well-matched by our sample (Figure~\ref{fig:stringplot}). Future work on orbital architecture of these systems combined with additional simulations tracking the inclination evolution of short-period planets are needed to determine or rule out secular interactions as an explanation.

In addition to these mechanisms mentioned above, other perturbative processes may also excite mutual inclinations, such as those related to stellar oblateness (e.g., \citealt{Spalding2016,Becker2020,Brefka2021,Faridani2024}) or stellar perturbers during disk phase (e.g., \citealt{Zanazzi2018}), although it is less obvious how they could correlate with stellar metallicity. For a more thorough discussion on the various mechanisms that could generate inclined planet orbits, we refer the readers to Section~5 of \cite{Millholland2021}.

\subsection{Caveats} 
\label{sec:discuss_caveats}
As mentioned in Section~\ref{sec:sample_selection}, our mutual inclination is estimated for the innermost two planets in each system assuming both planets transit the same stellar hemisphere. However, in practice, two transiting planets with identical impact parameters (inclinations) may transit across opposite hemispheres or have different ascending nodes, both resulting in non-zero mutual inclinations.

We argue that this caveat does not affect our main conclusion as following: (1) If two planets have inclinations of $i_1$ and $i_2$ (assuming $i_1 < i_2$ for simplicity as the result would be the same otherwise), $\Delta i$ will increase by $2 \times (90 - i_1)$ if the other planet transits in the opposite hemisphere. The median inclination of the innermost planets in our sample is $87.2^{\circ}$, which would lead to an increase of 5.6 degrees in $\Delta i$. (2) If two planets have different ascending nodes, the true mutual inclination could be arbitrarily large, but this would require a special viewing angle to see both planets transiting and is thus relatively rare. Furthermore, since systems with both high and low mutual inclinations have an equal probability of increasing their $\Delta i$ due to these effects, they should not impact our main conclusion that the metal-rich stars host systems with higher $\Delta i$ on average. However, they would alter the quantified mean and variance of $\Delta i$ for the two populations, rendering the reported values in this paper the lower limits.

\section{Conclusion and Future work}
\label{sec:conclusion_future}

In this work, we studied 89 multi-planet systems with the innermost planets within 10~days and reported a positive trend between stellar metallicity, [Fe/H], and mutual inclination, $\Delta i$, of the innermost two planets: short-period planets around metal-rich stars have larger and more diverse mutual inclinations. We considered the effect of stellar mass and age and concluded that metallicity is likely the dominant factor, though a connection to stellar mass cannot be entirely ruled out.

Future studies can further probe the origin of this correlation and test the theoretical explanations. For example, follow-up observations of high mutual inclination systems to confirm the presence or absence of additional non-transiting giant planets could be helpful, especially since the majority of the \textit{Kepler} systems in our sample lack RV follow-up. 
More in-depth studies taking into account \textit{Kepler} and TESS detection completeness and biases would yield better quantified intrinsic mutual inclination distributions. In addition, planetary population synthesis work often does not track inclinations or assumes coplanarity throughout. Therefore, we advocate for more theoretical and simulation works to target orbital inclinations and explore the origin and evolution of mutual inclinations in planetary systems.

\begin{acknowledgments}
X.\ Hua and S.\ X.\ Wang acknowledge support from NSFC grant 12273016 as well as Tsinghua Dushi Funding No.\ 53121000124.
Work by W. Zhu is supported by NSFC grant (No. 12173021 and 12133005).
X.\ Hua thanks Zhecheng Hu for his valuable suggestions on removing stellar flares. We thank Norm Murray and Sam Hadden for helpful discussions regarding theoretical interpretation.
We acknowledge the use of public TESS data from pipelines at the TESS Science Office and at the TESS Science Processing Operations Center.
Resources supporting this work were provided by the NASA High-End Computing (HEC) Program through the NASA Advanced Supercomputing (NAS) Division at Ames Research Center for the production of the SPOC data products.
This paper includes data collected with the TESS mission, obtained from the MAST data archive at the Space Telescope Science Institute (STScI). Funding for the TESS mission is provided by the NASA Explorer Program. STScI is operated by the Association of Universities for Research in Astronomy, Inc., under NASA contract NAS 5–26555.
All the TESS data used in this paper can be found in MAST: \dataset[doi: 10.17909/ak7h-3z66]{http://dx.doi.org/10.17909/ak7h-3z66}. 
This research has made use of the Exoplanet Follow-up Observation Program website, which is operated by the California Institute of Technology, under contract with the National Aeronautics and Space Administration under the Exoplanet Exploration Program (doi: 10.26134/ExoFOP3).
This research has made use of the NEA, which is operated by the California Institute of Technology, under contract with the National Aeronautics and Space Administration under the Exoplanet Exploration Program.
\end{acknowledgments}
%

\vspace{5mm}


\software{\code{lightkurve} \citep{lightkurve}, \code{Juliet} \citep{Juliet}, \code{dynesty} \citep{dynesty}, \code{PyMultiNest} \citep{pymultinest},  \code{UltraNest} \citep{Buchner2021}. }


\clearpage



\clearpage
\bibliography{2023reference}{}
\bibliographystyle{aasjournal}



\end{document}